\documentclass[a4paper,10pt]{article}
\usepackage{amsmath,amssymb,amsthm,latexsym,graphicx}
\pdfoutput=1

\newtheorem{lemma}{Lemma}
\newtheorem{corollary}{Corollary}
\newtheorem{prop}{Proposition}
\newtheorem{theorem}{Theorem}

\numberwithin{equation}{section}

\title{{\sc Some Remarks on T-copulas}}
\author{{\sc V. Frishling and D. G. Maher}\footnote{Group Market Risk, National Australia Bank.  24/255 George St, Sydney NSW 2000.  \;\;\;\;\; Corresponding author: $\langle$David.G.Maher@nab.com.au$\rangle$ }}

\begin{document}

\maketitle

\newcommand{\g}{\mathfrak{g}}
\newcommand{\kg}{\mathfrak{k}}
\newcommand{\tg}{\mathfrak{t}}
\newcommand{\ug}{\mathfrak{u}}
\newcommand{\p}{\mathfrak{p}}
\newcommand{\X}{\mathfrak{X}}
\newcommand{\af}{\mathfrak{a}}
\newcommand{\D}{\mathbf{D}}
\newcommand{\R}{\mathbb{R}}
\newcommand{\C}{\mathbb{C}}
\newcommand{\E}{\mathbb{E}}
\newcommand{\Z}{\mathbb{Z}}
\newcommand{\N}{\mathbb{N}}
\newcommand{\T}{\mathbb{T}}
\newcommand{\F}{\mathcal{F}}
\newcommand{\A}{\mathcal{A}}
\newcommand{\x}{\mathbf{x}}
\newcommand{\y}{\mathbf{y}}
\newcommand{\ad}{\mathrm{ad}}
\newcommand{\Ad}{\mathrm{Ad}}
\newcommand{\Exp}{\mathrm{Exp}}
\newcommand{\grad}{\mathrm{grad} \,}
\newcommand{\bbP}{\mathbb{P}}
\newcommand{\hp}{\mathfrak{h_p}}
\newcommand{\s}{\mathfrak{s}}
\newcommand{\h}{\mathfrak{h}}
\newcommand{\hf}{\mathfrak{h}}
\newcommand{\cosec}{\mathrm{cosec} \,}
\newcommand{\csch}{\mathrm{csch} \,}
\newcommand{\sgn}{\mathrm{sgn} \,}
\newcommand{\dd}{\, \mathrm{d}}
\newcommand{\Tr}{\, \mathrm{Tr}}

{\abstract We examine three methods of constructing correlated
Student-$t$ random variables.  Our motivation arises from
simulations that utilise heavy-tailed distributions for the purposes
of stress testing and economic capital calculations for financial
institutions.  We make several observations regarding the
suitability of the three methods for this purpose.\\

 }

\medskip
\medskip
\medskip
\medskip

\section{Introduction}

The use of heavy-tailed distributions for the purposes of stress
testing and economic capital calculations has gained attention
recently in an attempt to capture exposure to extreme events.\\

Among the various distributions available, the Student-t
distribution has gained popularity in these calculations for several
reasons (as opposed to, say, $\alpha$-stable distributions).  The
first is that for three or more degrees of freedom it possesses a
finite variance, and so can be calibrated to the variance of
observable data.  The second is that t-variables are
relatively easy and fast to generate for simulations.\\

However, one very desirable property that should be exhibited by any
calculation of economic capital is the ability to capture
concentrated risks.  Put simply, asset movements  - particularly
large movements  -  should be correlated.  Thus, it is necessary to
generate correlated t-variables.  A recent paper on this topic is
\cite{SL}  -  we refer the reader to this paper for the necessary
background on $t$-copulas, and the references contained therein.\\

In this paper we examine three t-copulas in this context, in
particular their properties regarding correlation and tail
correlation.

\medskip
\medskip

\section{T-Copulas}

Let $X, Y \sim N(0,1)$ with correlation $\rho(X,Y) = \rho$.\\

Typically, correlated Student-$t$ distributions with $n$ degrees of
freedom, $U$ and $V$, can be formed via the transformations:
\begin{equation}\label{SameChi}
U = X \sqrt{\frac{n}{C}} \; , \phantom{abcdefghij} V = Y
\sqrt{\frac{n}{C}}
\end{equation}
where $C$ is sampled from a chi-squared distribution with $n$
degrees of freedom\footnote{Formulations for Student-$t$
distributions with different degrees of freedom can be found in \cite{SL}.}\\

An alternative formulation is given by:
\begin{equation}\label{DiffChi}
U = X \sqrt{\frac{n}{C_1}} \; , \phantom{abcdefghij} V = Y
\sqrt{\frac{n}{C_2}}
\end{equation}
where $C_1$ and $C_2$ are independently sampled from a chi-squared
distribution with $n$ degrees of freedom.  This formulation is
suggested to be more desirable in \cite{SL} as it gives rise to a
product structure of the density function when $\rho = 0$.\\

However, we will show that this has a major impact on the
correlation, and in particular the resulting bivariate
distribution\footnote{We shall restrict our analysis in this paper
to the bivariate case only.} and tail correlation.\\

Another na\"{i}ve method of constructing correlated t-variables, $U$
and $V$, (assumed to have the same degrees of freedom) is the
following: take uncorrelated t-variables, $U$ and $W$, then put
\begin{equation}\label{CorrelT}
V = \rho \, U + \sqrt{1-\rho^2} \, W
\end{equation}

However, $V$ will not have a t-distribution as the sum of two
t-variables is not a t-variable.  Note that for three degrees of
freedom or more, the t-variable sums lie within the domain of
attraction of the Normal distribution.  However, since we are only
performing one sum, the tail of the distribution is still a power
law of order $n$.  Despite this, the resulting distribution does
posses some useful properties.\\

(\ref{SameChi}), (\ref{DiffChi}), and (\ref{CorrelT}) define the
three t-copulas that we will examine.  We refer to these t-copulas
as being generated by the {\it Same $\chi^2$, Independent $\chi^2$},
and {\it Correlated-t}, respectively.

\section{Independent $\chi^2$}

We will firstly examine the case of the Independent $\chi^2$
t-variables.  We now show that this construction has a major impact
on the correlation as follows:\\

Let $A = \sqrt{\frac{n}{C_1}}$ and $B = \sqrt{\frac{n}{C_2}}$. We
have,
\begin{align}
\rho(U,V) & = \frac{ \E(UV) - \E(U) \E(V) }{\sqrt{Var(U) Var(V)} }\\
& = \frac{ \E(XA \, YB) - \E(XA) \E(YB) }{\sqrt{Var(XA) Var(YB)} }\\
& = \frac{ \E(XY) \E(A) \E(B) - \E(X) \E(A) \E(Y) \E(B) }{\sqrt{Var(X) Var(Y) Var(A) Var(B)} } \tag{by independence} \\
& = \frac{ \E(XY) }{\sqrt{Var(X) Var(Y)}} \frac{ \E(A) \E(B) }{ \sqrt{ Var(A) Var(B)} } \tag{Since $\E(X) = \E(Y) = 0$} \\
& = \rho \, \frac{ \E(A) \E(B) }{\sqrt{ Var(A) Var(B)} }
\end{align}

Assuming that $A$ and $B$ have the same distribution, we have
\begin{align}
\rho(U,V) & = \rho \, \frac{ \E(A) \E(B) }{\sqrt{ Var(A) Var(B)} }\\
& = \rho \, \frac{ \E(A)^2 }{\E(A^2) }\\
& < \rho \tag{by Jensen}
\end{align}

In fact, the amount by which the correlation is reduced, namely
\begin{align}
\frac{ \E(A)^2 }{\E(A^2) } = \frac{ \E(n/C_1)^2 }{\E((n/C_1)^2) } =
\frac{ \E(1/C_1)^2 }{\E((1/C_1)^2) }
\end{align}
can be determined explicitly.  For the case where $C_1$ and $C_2$
have 3 degrees of freedom, this turns out to be $2/\pi \approx
0.6366$.\\

We now determine the amount by which the correlation is reduced by
explicitly.  We first begin with a calculation of the required
moments - we could not find a convenient reference, and record it
here for completeness:

\begin{lemma} The $n$th moment of the Inverse-Chi Distribution with
$\nu$ degrees of freedom is given by:
\begin{equation}
\E (Y^n) = \frac{\Gamma ((\nu - n)/2)}{\Gamma (\nu / 2)} 2^{n/2}
\end{equation}

\end{lemma}

{\bf Proof:} Let $f(x; \alpha, \beta)$ denote the Gamma
distribution, given by
\begin{equation}
f(x; \alpha, \beta) = \frac{\beta^\alpha}{\Gamma (\alpha)}
x^{\alpha-1} \exp(-x/\beta)
\end{equation}

If $\alpha = \nu/2$ and $\beta = 2$, then this is the chi-squared
distribution with $\nu$ degrees of freedom.\\

We wish to make the transformation $Y = 1/ \sqrt{X}$.  Since this is
a monotonic function, we use the transformation formula:
\begin{equation}
f_Y (y) = f_X \bigl( g^{-1} (y) \bigr) \Bigl| \frac{d}{dy} g^{-1}
(y) \Bigr|
\end{equation}

Thus,
\begin{align}
f_Y (y; \alpha, \beta) & = f_X \bigl( g^{-1} (y) \bigr) \Bigl|
\frac{d}{dy} g^{-1}
(y) \Bigr| \\
& = f_X \bigl( \frac{1}{y^2} \bigr) \Bigl| \frac{-2}{y^3} \Bigr| \\
& = \frac{\beta^\alpha}{\Gamma (\alpha)} (y^{-2})^{\alpha-1}
\exp(-(y^{-2})/\beta) \frac{2}{y^3} \\
& = \frac{\beta^\alpha}{\Gamma (\alpha)} y^{-2\alpha-1}
\exp(-1/(\beta y^2))\\
\end{align}

We now derive the formula for the $n$-th moment of $Y$. Firstly,
note that
\begin{equation}
\int_0^\infty y^{-2\alpha-1} \exp(-1/(\beta y^2)) dy = \frac{\Gamma
(\alpha)}{\beta^\alpha}
\end{equation}
and we have:
\begin{align}
\E (Y^n) & = \int_0^\infty y^n \frac{\beta^\alpha}{\Gamma (\alpha)}
y^{-2\alpha-1} \exp(-1/(\beta y^2)) \\
& =  \frac{\beta^\alpha}{\Gamma (\alpha)} \int_0^\infty
y^{-2(\alpha - n/2) -1} \exp(-1/(\beta y^2)) \\
& =  \frac{\beta^\alpha}{\Gamma (\alpha)} \frac{\Gamma
(\alpha - n/2)}{\beta^{\alpha-n/2}}\\
& =  \frac{\Gamma (\alpha - n/2)}{\Gamma (\alpha)} \beta^{n/2}
\end{align}

{\bf Remark:} Note that the moments for this distribution will not
be defined when $\alpha - n/2$ is a negative integer.\\

Consider now the case when $\alpha = \nu/2$ and $\beta = 2$:
\begin{equation}
\E (Y^n) = \frac{\Gamma ((\nu - n)/2)}{\Gamma (\nu / 2)} 2^{n/2}
\end{equation}

as required. $\phantom{abcde} \square$\\

\begin{prop}  The factor by which the correlation is reduced by is
given by
\begin{equation}
\frac{ [\E (Y)]^2 }{ \E (Y^2)} = \Biggl( \frac{\Gamma ((\nu -
1)/2)}{\Gamma (\nu / 2)} \Biggr)^2 \times \Biggl( \frac{\nu - 2}{2}
\Biggr)
\end{equation}
Furthermore, we have for $\nu = 3$,
\begin{equation}
\frac{ [\E (Y)]^2 }{ \E (Y^2)} = \frac{2}{\pi} \approx 0.6366
\end{equation}
and for large $\nu$ we have
\begin{equation}
\frac{ [\E (Y)]^2 }{ \E (Y^2)} = \frac{\nu - 2}{\nu - 1} \rightarrow
1 \; \; \text{as} \; \; \nu \rightarrow \infty
\end{equation}
\end{prop}

{\bf Proof:}  Let us first consider the case $\nu = 3$:
\begin{equation}
\E (Y^n) = \frac{\Gamma ((3 - n)/2)}{\Gamma (3 / 2)} 2^{n/2}
\end{equation}

Hence,
\begin{equation}
\E (Y) = \frac{\Gamma (1)}{\Gamma (3 / 2)} 2^{1/2} =
\frac{1}{\sqrt{\pi} / 2} 2^{1/2} = \frac{2\sqrt{2}}{\sqrt{\pi}}
\end{equation}

and
\begin{equation}
\E (Y^2) = \frac{\Gamma (1/2)}{\Gamma (3 / 2)} 2 =
\frac{\sqrt{\pi}}{\sqrt{\pi} / 2} 2 = 4
\end{equation}

Thus we have,
\begin{equation}
\frac{ [\E (Y)]^2 }{ \E (Y^2)} = \Biggl(
\frac{2\sqrt{2}}{\sqrt{\pi}} \Biggr)^2 \times \Biggl( \frac{1}{4}
\Biggr) = \frac{2}{\pi} \approx 0.6366
\end{equation}

For general degrees of freedom we have
\begin{align}
\frac{ [\E (Y)]^2 }{ \E (Y^2)} & = \Biggl( \frac{\Gamma ((\nu -
1)/2)}{\Gamma (\nu / 2)} 2^{1/2} \Biggr)^2 \times \Biggl(
\frac{\Gamma (\nu / 2)}{\Gamma ((\nu - 2)/2)} 2^{-2/2} \Biggr) \\
& = \Gamma^2 ((\nu - 1)/2) \times \Bigl( \Gamma ((\nu - 2)/2) \Gamma
(\nu / 2) \Bigr)^{-1}\\
& = \Gamma^2 ((\nu - 1)/2) \times \Bigl( 2/(\nu - 2) \times \Gamma^2
(\nu/2) \Bigr)^{-1}\\
& = \Biggl( \frac{\Gamma ((\nu - 1)/2)}{\Gamma (\nu / 2)} \Biggr)^2
\times \Biggl( \frac{\nu - 2}{2} \Biggr)\label{corrRed}
\end{align}

{\bf Remark:} Compare (\ref{corrRed}) with the expression given in
\cite{SL}, section 4.1, which is very similar, except that they
(erroneously) give the square-root of this expression.\\

Using the properties of Beta functions and Stirling's formula, we
have that
\begin{align}
\frac{\Gamma ((\nu - 1)/2)}{\Gamma (\nu / 2)} & =
\frac{B(\tfrac{1}{2}, \tfrac{\nu - 1}{2})}{\sqrt{\pi}}\\
& = \sqrt{\frac{2}{\nu - 1}}
\end{align}
and thus
\begin{equation}
\frac{ [\E (Y)]^2 }{ \E (Y^2)} = \frac{\nu - 2}{\nu - 1}
\end{equation}
as required. $\phantom{abcde} \square$\\

To explicitly compute (\ref{corrRed}) for a given value of $\nu$, we
need to consider odd and even cases.  We table here the following
values:


\begin{center}
    \begin{tabular}{ | c | c |}
    \hline
    Degrees of Freedom & Reduction of Correlation \\ \hline
    3 & 0.6366 \\ \hline
    4 & 0.7854 \\ \hline
    5 & 0.8488 \\ \hline
    6 & 0.8836 \\ \hline
    7 & 0.9054 \\ \hline
    8 & 0.9204 \\ \hline
    9 & 0.9313 \\ \hline
    10 & 0.9396 \\ \hline
    20 & 0.9726 \\ \hline
    50 & 0.9896 \\ \hline
    100 & 0.9949 \\ \hline
    \end{tabular}
\end{center}



\section{Empirical Distributions and Copulas}

We now provide empirical results for each of our three t-copulas. We
simulated 1,000,000 observations using each of the three methods,
and have provided below graph of the pdf's of the distribution and
the copulas.  We have only considered here the case of the
t-distribution with three degrees of freedom, and a base correlation
of 0.9. (the graphs presented in this section are based on the first
5,000 observations)\\

\begin{figure}[htp]
\centering
\includegraphics{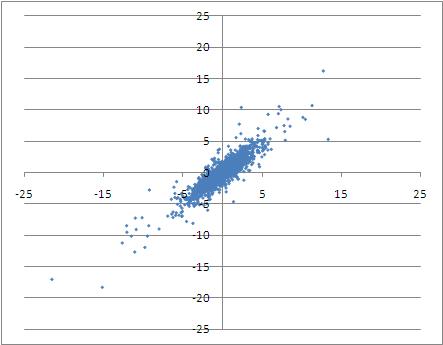}
\caption{Same $\chi^2$} \label{fig:SameChiPDF}
\end{figure}

\begin{figure}[htp]
\centering
\includegraphics{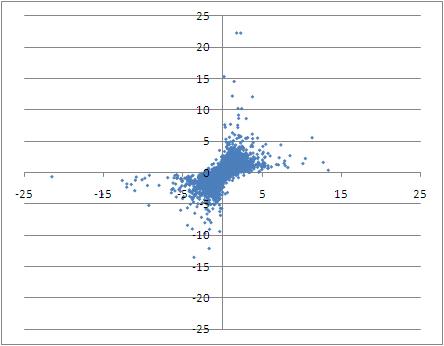}
\caption{Independent $\chi^2$} \label{fig:IndepChiPDF}
\end{figure}

\begin{figure}[htp]
\centering
\includegraphics{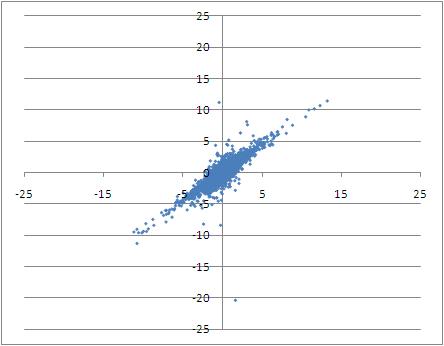}
\caption{Correlated-$t$} \label{fig:CorrelT}
\end{figure}

\newpage

As can be clearly seen, the pdf's of the Same $\chi^2$ and
Correlated-$t$ are elliptical, but the pdf of the Independent
$\chi^2$ is quite splayed out.\\

We have also constructed graphs of their copulas (below).  Despite
being slightly more ``fatter'', the copula of the Independent
$\chi^2$ appear little different.

\begin{figure}[htp]
\centering
\includegraphics{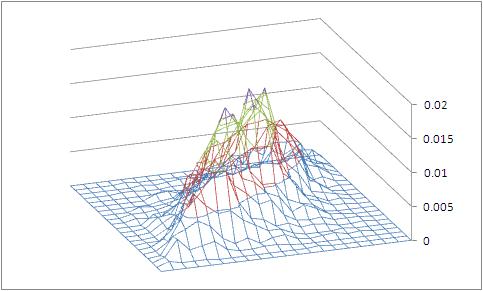}
\caption{Same $\chi^2$ Copula Density} \label{fig:SameChiCopula}
\end{figure}

\begin{figure}[htp]
\centering
\includegraphics{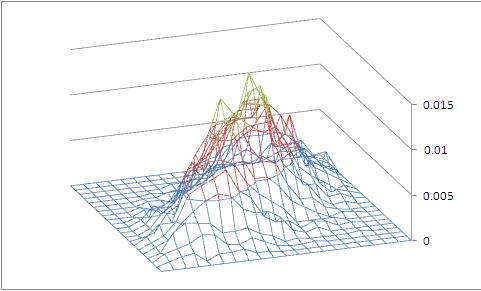}
\caption{Independent $\chi^2$ Copula Density}
\label{fig:IndepChiCopula}
\end{figure}

\begin{figure}[htp]
\centering
\includegraphics{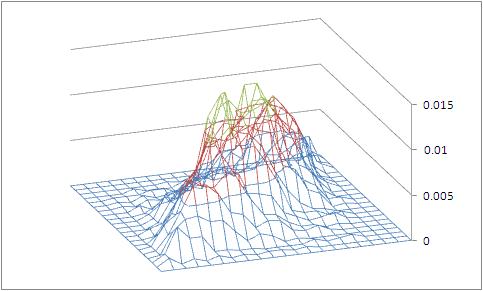}
\caption{Correlated-$t$ Copula Density} \label{fig:CorrelT}
\end{figure}

\newpage

\section{Tail Correlation}

We firstly prove a general result concerning the tail correlation.
Recall that the tail correlation is the quantity
\begin{equation}
correl(U,V \, | \, U > \gamma)
\end{equation}
where $\gamma$ is the number of standard deviations into the tail.\\

Consider two random variables, $Z$ and $X$, having correlation
$\rho$, formed by the sum:
\begin{equation}
Z = \rho X + \sqrt{1 - \rho^2} Y
\end{equation}
where $X$ and $Y$ are independent random variables.

\begin{theorem}  The tail correlation of $X$ and $Z$ is given by
\begin{equation}
correl(X,Z | X > \mu) = \frac{1}{\sqrt{1 + \frac{K'}{V}}}
\end{equation}
where
\begin{equation}
V = Var [X | X > \mu] \phantom{abc} \text{and} \phantom{abc} K' =
Var[Y]/\rho^2
\end{equation}

Furthermore, suppose the tails of $X$ and $Y$ have a power law, then
\begin{equation}
correl(X,Z | X > \mu) \; \rightarrow 1 \phantom{abc} \text{as}
\phantom{abc} \mu \rightarrow \infty
\end{equation}
and $X$ and $Y$ are Normal variables, then
\begin{equation}
correl(X,Z | X > \mu) \; \rightarrow 0 \phantom{abc} \text{as}
\phantom{abc} \mu \rightarrow \infty
\end{equation}
\end{theorem}

The corollary for the $t$-distribution follows from our proof of
this theorem:
\begin{corollary}  Let $X$ and $Y$ be independent $t$-variables with $\nu$ degrees of freedom.  Then the tail variance of $X$ is given by
\begin{equation}
V = Var [X | X > \mu] = \mu^2 \frac{\nu}{(\nu-1)^2(\nu-2)}
\end{equation}
and therefore
\begin{equation}
correl(X,Z | X > \mu) = \frac{1}{\sqrt{1 + \frac{K'
(\nu-1)^2(\nu-2)}{\mu^2 \nu}}}
\end{equation}
\end{corollary}

{\bf Proof:}  Consider the conditional correlation, given $X > \mu$.
Then we have
\begin{align*}
\E [XZ | X > \mu] & = \rho \E[X^2 | X > \mu] + \sqrt{1 - \rho^2}
\E[XY | X > \mu]\\
& = \rho \E[X^2 | X > \mu]
\end{align*}

\begin{equation}
\E [X | X > \mu] \E [Z | X > \mu] = \rho \E [X | X > \mu]^2
\end{equation}

\begin{equation}
Var [X | X > \mu] = \E [X^2 | X > \mu] - \E [X | X > \mu]^2 = V
\phantom{abcde} \text{(say)}
\end{equation}

\begin{align*}
VaR [Z | X > \mu] & = \rho^2 Var[X | X > \mu] + (1 - \rho^2)
Var[Y | X > \mu]\\
& = \rho^2 V + (1 - \rho^2) Var[Y | X > \mu]\\
& = \rho^2 V + K\tag{say}
\end{align*}

and
\begin{equation}
correl(XZ) = \frac{\rho V}{\sqrt{V} \sqrt{\rho^2 V + K}} =
\frac{1}{\sqrt{1 + \frac{K'}{V}}}
\end{equation}

so the behaviour of the tail correlation depends on the tail
variance $V$.  We firstly examine power law tails, then the Normal
distribution.\\

{\bf Power law tails:}\\

Let the tail of a distribution (density) be $f(x) = C x^{-n}$  The
conditional distribution is:
\begin{equation}
F(x | X > \mu) = \frac{Pr(\mu < X < x)}{Pr(X > \mu)} =
\frac{\mu^{-n+1} - x^{-n+1}}{\mu^{-n+1}}
\end{equation}
and the conditional density is:
\begin{equation}
f(x | X > \mu) = \frac{(n-1) x^{-n}}{\mu^{-n+1}}
\end{equation}

Thus, the conditional variance (that is, $V$) is given by
\begin{align*}
VaR [X | X > \mu] & = \E [X^2 | X > \mu] - \E [X | X > \mu]^2\\
& = \int_\mu^\infty \frac{(n-1) x^{-n+2}}{\mu^{-n+1}} dx - \biggl(
\int_\mu^\infty \frac{(n-1) x^{-n+1}}{\mu^{-n+1}} dx \biggr)^2\\
& = \frac{(n-1) \mu^{-n+3}}{(n-3) \mu^{-n+1}} - \biggl( \frac{(n-1)
\mu^{-n+2}}{(n-2) \mu^{-n+1}} \biggr)^2\\
& = \frac{(n-1)}{(n-3)}\mu^{2} - \frac{(n-1)^2}{(n-2)^2} \mu^{2} \\
& = \frac{n-1}{(n-2)^2(n-3)} \mu^2 \; \rightarrow \infty
\phantom{abc} \text{as} \phantom{abc} \mu \rightarrow \infty
\end{align*}

and thus
\begin{equation}
correl(XZ | X > \mu) = \frac{1}{\sqrt{1 + \frac{K'}{V}}} \;
\rightarrow 1 \phantom{abc} \text{as} \phantom{abc} \mu \rightarrow
\infty
\end{equation}

{\bf Normal Distribution:}\\

The conditional distribution for the Normal distribution is:
\begin{equation}
F(x | X > \mu) = \frac{Pr(\mu < X < x)}{Pr(X > \mu)} = \frac{N(x) -
N(\mu)}{1 - N(\mu)}
\end{equation}
and the conditional density is:
\begin{equation}
f(x | X > \mu) = \frac{n(x)}{1 - N(\mu)}
\end{equation}

Now
\begin{equation}
\int_\mu^\infty x^2 n(x) dx = \mu n(\mu) + 1 - N(\mu)
\end{equation}
and
\begin{equation}
\int_\mu^\infty x n(x) dx = n(\mu)
\end{equation}

Thus, the conditional variance (that is, $V$) is given by
\begin{align*}
VaR [X | X > \mu] & = \E [X^2 | X > \mu] - \E [X | X > \mu]^2\\
& = \frac{1}{1-N(\mu)} \biggl[\int_\mu^\infty x^2 n(x) dx -
\frac{1}{1-N(\mu)} \biggl(\int_\mu^\infty x n(x) dx \biggr)^2
\biggr]\\
& = \frac{\mu n(\mu) + 1 - N(\mu)}{1 - N(\mu)} - \frac{n(\mu)^2}{[1
- N(\mu)]^2}\\
& = 1+ \frac{\mu n(\mu)}{1 - N(\mu)} - \frac{n(\mu)^2}{[1 -
N(\mu)]^2}
\end{align*}

It can be shown using l'h\^{o}pital's rule that this expression
tends to zero as $\mu$ tends to infinity.  Thus we have,
\begin{equation}
correl(XZ) = \frac{1}{\sqrt{1 + \frac{K'}{V}}} \; \rightarrow 0
\phantom{abc} \text{as} \phantom{abc} \mu \rightarrow 0
\end{equation}

which concludes the proof. $\phantom{abc} \square$

\section{Tail Correlations: Empirical Results}

We now examine numerical calculations of the tail correlations for
each of our three t-copulas. That is, we estimate the quantity
\begin{equation}
correl(U,V \, | \, U > \gamma)
\end{equation}
where $\gamma$ is the number of standard deviations into the tail.
We record the tail correlation for each of our three t-copulas  -
using the same base 0.9 correlation and three degrees of freedom  -
for the given standard deviation below:



\begin{center}
    \begin{tabular}{ | c || c | c | c |}
    \hline
    Tail St. Dev. & \multicolumn{3}{|c|}{Tail Correlation} \\
    \hline
    $\gamma$ & Same $\chi^2$ & Different $\chi^2$ & Correlated-t \\ \hline
    2 & 0.8273 & 0.0430 & 0.9314 \\ \hline
    3 & 0.8255 & 0.0124 & 0.9562 \\ \hline
    4 & 0.8210 & -0.0046 & 0.9723 \\ \hline
    5 & 0.8168 & -0.0046 & 0.9800 \\ \hline
    6 & 0.8113 & -0.0011 & 0.9839 \\ \hline
    7 & 0.8074 & -0.0021 & 0.9880 \\ \hline
    8 & 0.8049 & -0.0148 & 0.9899 \\ \hline
    9 & 0.8036 & -0.0199 & 0.9912 \\ \hline
    10 & 0.8044 & -0.0128 & 0.9927 \\ \hline
    11 & 0.8033 & -0.0024 & 0.9934 \\ \hline
    12 & 0.8003 & -0.0031 & 0.9942 \\ \hline
    13 & 0.8015 & -0.0094 & 0.9941 \\ \hline
    14 & 0.8053 & -0.0132 & 0.9942 \\ \hline
    15 & 0.8023 & -0.0120 & 0.9938 \\ \hline
    16 & 0.8027 & -0.0132 & 0.9942 \\ \hline
    17 & 0.8016 & -0.0206 & 0.9938 \\ \hline
    18 & 0.7991 & -0.0247 & 0.9945 \\ \hline
    19 & 0.7983 & -0.0390 & 0.9944 \\ \hline
    20 & 0.7927 & -0.0338 & 0.9949 \\ \hline
    \end{tabular}
\end{center}

\medskip
\medskip

As can be seen, the tail correlation for the t-distribution using
the Same $\chi^2$ remains high, and decays slightly into the tail to
approximately 0.8.  Unsurprisingly for the t-distribution using
Different $\chi^2$, the tail correlation is quite low, and even
becomes slightly negative into the tail.  This is a feature of the
tail, and can be seen in Figure 7 below, which is a graph of all
observations where the first variable.  As predicted by Theorem 1,
the tail correlation of the Correlated-t {\it increases} into the
tail.\\

\begin{figure}[htp]
\centering
\includegraphics{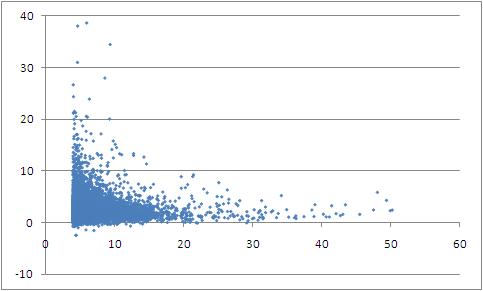}
\caption{Independent $\chi^2$ Tail} \label{fig:IndepChiTail}
\end{figure}

\medskip
\medskip

\newpage

We also give the number of tail observations (out of 1,000,000) for
each of our three t-copulas, again using the same base 0.9
correlation and three degrees of freedom. That is, the quantity
\begin{equation}
\sharp(U,V \, | \, U > \gamma , \, V > \gamma)
\end{equation}

\begin{center}
    \begin{tabular}{ | c || c | c | c |}
    \hline
    Tail St. Dev. & \multicolumn{3}{|c|}{Tail Observations} \\
    \hline
    $\gamma$ & Same $\chi^2$ & Different $\chi^2$ & Correlated-t \\ \hline
    2 & 49925 & 24825 & 53046 \\ \hline
    3 & 19918 & 5558 & 21880 \\ \hline
    4 & 9534 & 1524 & 10472 \\ \hline
    5 & 5199 & 451 & 5706 \\ \hline
    6 & 3099 & 165 & 3421 \\ \hline
    7 & 2039 & 71 & 2214 \\ \hline
    8 & 1417 & 37 & 1530 \\ \hline
    9 & 998 & 22 & 1088 \\ \hline
    10 & 726 & 9 & 800 \\ \hline
    11 & 556 & 5 & 606 \\ \hline
    12 & 439 & 3 & 476 \\ \hline
    13 & 334 & 0 & 383 \\ \hline
    14 & 258 & 0 & 305 \\ \hline
    15 & 208 & 0 & 255 \\ \hline
    16 & 166 & 0 & 204 \\ \hline
    17 & 143 & 0 & 171 \\ \hline
    18 & 118 & 0 & 147 \\ \hline
    19 & 97 & 0 & 127 \\ \hline
    20 & 88 & 0 & 113 \\ \hline
    \end{tabular}
\end{center}

\medskip
\medskip



\section{Summary and Conclusions}

We have examined the t-copulas for the purposes of stress testing
and economic capital calculations.  It appears that using correlated
t-variables generated by using different $\chi^2$ is not appropriate
as the correlation is affected (and essentially destroyed) by the
construction.  Whilst the Correlated-t is not a true t-distribution
with the desired degrees of freedom, the distribution is still heavy
tailed, and has the desired properties regarding correlation and in
particular tail correlation.

\medskip
\medskip
\medskip

\medskip
\medskip
\medskip



\begin{thebibliography}{12}
\small

\bibitem[SL]{SL}
{\sc Shaw, W.T. and Lee, K.T.A.}  {\it Copula Methods vs Canonical
Multivariate Distributions: the multivariate Student T distribution
with general degrees of freedom,} [2007], Preprint available at:
http://www.mth.kcl.ac.uk/$\sim$shaww/web\_page/papers/MultiStudentc.pdf







\end{thebibliography}
\end{document}